\documentclass[aps,floatfix,prd,twocolumn,showpacs,groupedaddress]{revtex4}

\usepackage[cmex10]{amsmath}
\usepackage{amssymb}
\usepackage{latexsym}
\usepackage{wrapfig}
\usepackage{graphicx}
\usepackage[tight,footnotesize]{subfigure}
\usepackage{multirow}

\usepackage{deleq}
\usepackage{dcolumn}
\usepackage{color}
\usepackage{ulem}

\newcommand{\bma}[1]{\mbox{\boldmath${#1}\/$}}
\newcommand{\Cdot}{\bma{\cdot}}
\newcommand{\Nabla}{\bma{\nabla}}

\begin{document}

\title[Inflight magnetic characterization of LISA Pathfinder]{Inflight
  magnetic   characterization  of   the  test   masses   onboard  LISA
  Pathfinder}

\author{Marc Diaz-Aguil\'o$^{1,2}$} 
\email[]{marc.diaz.aguilo@fa.upc.edu}
\author{Enrique Garc\'\i a--Berro$^{1,2}$}
\affiliation{$^1$Departament de F\'\i sica Aplicada, 
                 Universitat Polit\`ecnica de Catalunya, 
                 c/Esteve Terrades, 5, 
                 08860 Castelldefels, 
                 Spain\\
             $^2$Institut d'Estudis Espacials de Catalunya,
                 c/Gran Capit\`a 2--4, 
                 Edif. Nexus 104, 
                 08034 Barcelona, 
                 Spain}
\author{Alberto Lobo$^{3,2}$}
\affiliation{$^3$Institut de Ci\`encies de l'Espai, CSIC, 
                 Campus UAB, Facultat de Ci\`encies, Torre C-5, 
                 08193 Bellaterra, 
                 Spain}

\date{\today}

\begin{abstract}
LISA  Pathfinder  is a  science  and  technology  demonstrator of  the
European Space  Agency within the  framework of its LISA  mission, the
latter  aiming   to  be  the  first   space-borne  gravitational  wave
observatory.   The payload of  LISA Pathfinder  is the  so-called LISA
Technology   Package,   which   is   designed  to   measure   relative
accelerations  between two  test  masses in  nominal  free fall.   The
diagnostics subsystem consists of several modules, one of which is the
magnetic diagnostics unit. Its main  function is the assessment of the
differential  acceleration  noise  between  the  test  masses  due  to
magnetic  effects.  This subsystem  is composed  of two  onboard coils
intended to produce controlled magnetic  fields at the location of the
test masses.  These magnetic  fields couple with the remanent magnetic
moment and susceptibility  and produce forces and torques  on the test
masses.   These, in  turn, produce  kinematic excursions  of  the test
masses which are sensed by  the onboard interferometer.  We prove that
adequately processing these excursions, the magnetic properties of the
test   masses  can  be   estimated  using   classical  multi-parameter
estimation  techniques.   Moreover, we  show  that special  processing
procedures to minimize the effect of the multi channel cross-talks are
needed.  Finally, we demonstrate that  the quality of our estimates is
frequency dependent.  We also  suggest that using a multiple frequency
experiment the global estimate can be  obtained in such a way that the
results of  the magnetic experiment are more  reliable. Finally, using
our procedure we compute the the contribution of the magnetic noise to
the total proof-mass acceleration noise.
\keywords{LISA Pathfinder \and  magnetic characteristics \and on-board
  instrumentation \and space borne and space-research instruments \and
  parameter estimation}
\end{abstract}

\pacs{04.80.Nn, 04.30.-w, 07.87.+v, 06.30.Ka, 07.05.Fb}

\maketitle


\section{Introduction}
\label{chap.1}

LISA  Pathfinder  (LPF)  is  a  science  and  technology  demonstrator
programmed by the European Space  Agency (ESA) within its LISA mission
activities~\cite{bib:LISA}.  LISA (Laser Interferometer Space Antenna)
is  a joint ESA-NASA  mission which  will be  the first  low frequency
(milli-Hz) gravitational wave detector, and also the first space-borne
gravitational  wave  observatory.   The   payload  of  LPF,  the  LISA
Technology  Package (LTP),  will be  the highest  sensitivity geodesic
explorer  flown to  date.  The  LTP  is designed  to measure  relative
accelerations between  two test masses in nominal  free fall (geodesic
motion) with a differential acceleration noise budget
\begin{equation}
 S^{1/2}_{\delta a, LPF}(\omega) \leq 3 \times 10^{-14} \left[ 1 +
 \left(\frac{\omega/2\pi}{3\;
\rm{mHz}}\right)^2 \right]  \frac{\rm{m\; s}^{-2}}{\sqrt{\rm Hz}}
\label{eq.0}
\end{equation}
in  the frequency  band between  1\,mHz  and 30\,mHz~\cite{bib:trento,
bib:LPFmission}.

Magnetic noise in  the LTP is allowed to be  a significant fraction of
the      total     mission      acceleration     noise:      up     to
$1.2\times10^{-14}$\,m\,s$^{-2}$\,Hz$^{-1/2}$    is   apportioned   to
magnetic    effects,    i.e.,    40\,\%    of   the    total    noise,
3$\times$10$^{-14}$\,m\,s$^{-2}$\,Hz$^{-1/2}$,  see  Eq.~(\ref{eq.0}).
This   noise   occurs   because   the   residual   magnetization   and
susceptibility of  the test masses couple to  the surrounding magnetic
field, giving rise to a fluctuating force which is given by:
\begin{equation}
 {\bf \delta F} = \left\langle\left[\left({\bf M} +
 \frac{\chi}{\mu_0}\,{\bf B}\right)\Cdot \delta \Nabla\right]{\bf B}+
 \frac{\chi}{\mu_0}\,\left[\delta{\bf B}\Cdot\Nabla\right]{\bf B}
 \right\rangle V
 \label{eq.2}
\end{equation}

\noindent in each  of the test masses.  In this  expression {\bf B} is
the  magnetic field  in  the test  mass,~$\chi$  and {\bf  M} are  its
magnetic  susceptibility  and  residual  density of  magnetic  moment,
respectively, and $V\/$ is the volume of the test mass, $\mu_0$ is the
vacuum            magnetic            constant,            $4\pi\times
10^{-7}$\,m\,kg\,s$^{-2}$\,A$^{-2}$,  and   $\langle  \cdots  \rangle$
indicates test mass volume  average of the enclosed quantity. Finally,
$\delta{\bf B}$ represents the  fluctuation of the magnetic field, and
$\delta\Nabla$     stands    for     the     fluctuation    of     the
gradient~\cite{bib:ntcs}.   Quantitative  assessment  of the  magnetic
noise in the LTP, i.e. evaluation of Eq.~(\ref{eq.2}) clearly requires
a real-time monitoring of the magnetic field and an accurate knowledge
of the magnetic characteristics of the test masses.

The determination  of the magnetic characteristics of  the test masses
(remanent magnetic  moment and susceptibility) must be  done in flight
because their  magnetic properties may  change due to  launch stresses
and  other  circumstances.  This  will  be  done injecting  controlled
sinusoidal magnetic  fields at  the positions of  the test  masses and
appropriately  processing  the  resulting  kinematics, which  will  be
obtained from  the readings  delivered by the  onboard interferometer.
Although the basic  design of the magnetic experiment  is well settled
\cite{bib:myPaper3}, due to the high complexity of the LTP experiment,
more  in-depth  analyses  based  on  a  more  realistic  modeling  are
necessary to  assess its feasibility and performance.   The purpose of
this paper is,  precisely, to fill this gap.   In particular, we model
in a realistic  way the kinematics of the test  masses and we evaluate
the  expected quality  of the  estimates  of the  magnetic moment  and
susceptibility.   Specifically, we take  into account  several effects
--- like  the  cross-talks  between   some  of  the  channels  of  the
instrument, or the frequency-dependent parameters of the control loops
governing the dynamics  of the test masses ---  that previous analyses
disregarded.  All these effects depend on the frequency used to excite
the test masses.  Hence, the  quality of the estimates of the magnetic
data  depends  sensitively  on  the excitation  frequency,  since  the
satellite  does not  behave  equally across  the complete  measurement
bandwidth. Therefore, it is important  to determine the quality of the
estimates   across  the  complete   measurement  bandwidth,   and  the
frequencies that deliver the best estimate of the magnetic parameters.

The paper  is organized as  follows.  In Sect.~\ref{chap.2} we  give a
brief description of the  magnetic experiment intended to estimate the
magnetic properties  of the test masses.   Then, in Sect.~\ref{chap.3}
we briefly present  the dynamical model of the  satellite.  It follows
Sect.~\ref{chap.4},  where we  discuss  the estimation  model and  the
estimation  procedures used  in this  work.  In  Sect.~\ref{chap.5} we
present the sensitivity of this model to different hardware systems of
the  satellite and  in  Sect.~\ref{chap.6} we  evaluate the  frequency
dependence  of the experiment,  and we  optimize its  performance with
respect to the excitation frequency.  Sect.~\ref{chap.7} is devoted to
analyze  the  robustness  of  our  results.   Sect.~\ref{chap.MagCont}
determines the  expected accuracy of the magnetic  contribution to the
total proof-mass acceleration noise. Finally, in Sect.~\ref{chap.8} we
summarize  our  main findings,  we  discuss  the  significance of  our
results, and we draw our conclusions.


\section{Experiment overview}
\label{chap.2}

\begin{figure}[!t]
 \centering
 \includegraphics[width=0.8\columnwidth]{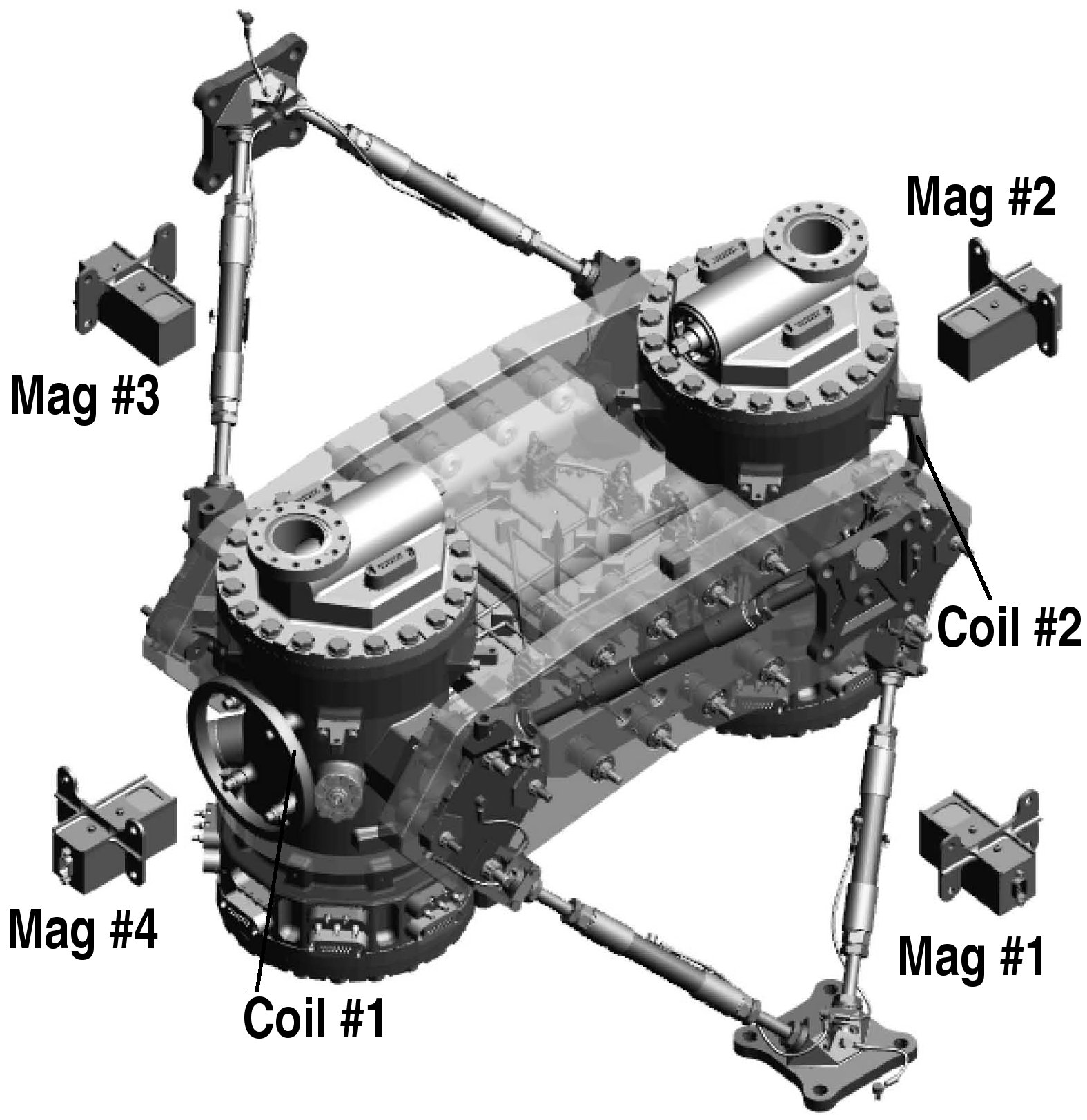}
  \caption{A  schematic view of  the payload  of LISA  Pathfinder, the
    LTP.  The Inertial Sensors (two vertical towers) host the two test
    masses.   The  four floating  boxes  correspond  to the  tri-axial
    fluxgate  magnetometers, and  the two  induction coils  are placed
    next  to  each of  the  test masses.   The  optical  bench of  the
    interferometer  is located  on  the horizontal  plane between  the
    Inertial Sensors.}
 \label{fig:LTP}
\end{figure}

The two test masses are located  at the center of each inertial sensor
--- the two towers  in Fig.~\ref{fig:LTP} --- and are  the end mirrors
of  the  Optical  Metrology  System,  that senses  the  positions  and
attitudes of the test masses.  The optical bench of the interferometer
can be seen  in Fig.~\ref{fig:LTP} as well.  In fact,  one of the test
masses is the reference free  floating body to perform the translation
and  attitude  control  of   the  spacecraft.   The  $x$-axis  of  the
experiment is the axis connecting  the two test masses centers, and it
goes from test mass 1 to test mass 2.  The $z$-axis points towards the
solar panel (parallel to the two inertial sensor towers and upwards in
Fig.~\ref{fig:LTP}) and, finally, the $y$-axis closes the right-handed
Cartesian coordinate system.  The test  masses are made of an alloy of
Pt (27\%) and Au (73\%), their dimensions are $46\times46\times 46$~mm
and  their  weight  is  1.95~kg.   To  comply  with  the  top  science
requirements, the  test masses must have certain  properties.  For the
purpose of the present work  the two most important properties are the
remanent  magnetic  moment   and  the  susceptibility.   The  remanent
magnetic moment must  be $|{\bf m}|<2.0\times 10^{-8}$~A~m$^2$.  Since
the volume of  the test masses is $V =  0.046^3$~m$^3$, the density of
magnetic moment must be then $|{\bf M}|<9.451 \cdot 10^{-4}$~A/m.  The
susceptibility  of the  test mass  can  be suitably  represented by  a
complex  number,  $\chi  =  \chi_{\rm  o} +  i  \chi_{\rm  e}$,  where
$\chi_{\rm o}$ is  its real component and $\chi_{\rm  e}(\omega)$ is a
frequency-dependent imaginary  term which is due to  the eddy currents
on the  test mass \cite{bib:trento,  bib:LPFmission}.  The requirement
on  the  value  of  the  real  component  is  $\chi_{\rm  o}<2.5\times
10^{-5}$.

As  mentioned,  to  measure  the  remanent  magnetic  moment  and  the
susceptibility of the test masses  a controlled magnetic field will be
injected  at the  position of  the test  masses.  This  magnetic field
produces forces  and torques which  excite the kinematics of  the test
masses.   Studying  the  motion  of  the  test  masses,  namely  their
displacement and rotation, allows  to estimate the three components of
the magnetic moment and the susceptibilities of the test masses.

\subsection{The injected magnetic fields}

The  magnetic  field  at the  position  of  the  test masses  will  be
generated  by the  injection  of sinusoidal  currents  to the  onboard
coils. These  onboard coils  are placed next  to each of  the inertial
sensors  towers,  see  again  Fig.~\ref{fig:LTP}.   The  two  circular
induction coils are made of  a titanium alloy (${\rm Ti_6Al_4V}$), and
have        $N=$        2\,400        windings        of        radius
$r=56.5$~mm~\cite{bib:LPFmission}. They  are placed 85.5  mm away from
the center of the respective test mass.  The onboard coils are aligned
with the $x$-axis of the  test masses, thus, the magnetic field within
the volume of the test masses has axial symmetry.  Given a current fed
to the  coils $I(t)=I_0\sin\omega_0  t$, the resulting  magnetic field
(and its gradient) will oscillate  at the same frequency. Therefore we
write,

\begin{eqnarray}
 \textbf{B}_{\rm app} & = & {\rm Re}\left\{ \textbf{B}_0\, 
 i e^{-i \omega_0 t}  \right\} =  \textbf{B}_0\,\sin\omega_0 t 
 \label{eq.appfield} \\
 \Nabla\textbf{B}_{\rm app} & = & {\rm Re}\left\{ \Nabla\textbf{B}_0\, 
 i e^{-i \omega_0 t}  \right\}  = \Nabla\textbf{B}_0\,\sin\omega_0 t 
 \label{eq.appfield2}
\end{eqnarray}

The field  produced by  the coils at  the center  of the test  mass is
$4.47~\mu$T, whereas the maximum environmental magnetic field expected
during science operation is less than 100\,nT.  On the other hand, the
magnetic field  gradient along the  $x$-axis produced by the  coils is
109.2~$\mu$T/m, while the maximum  magnetic field gradient required by
the mission  science specification is $-5~\mu$T/m. Therefore,  it is a
safe assumption  to neglect the effects of  the environmental magnetic
field with respect to the applied field by the coils. Thus, the forces
and   torques   exerted  on   the   test   masses   are  computed   as
\cite{bib:jackson}:
\begin{equation}
 {\bf F} = \left\langle\left[\left({\bf M} + 
 {\rm Re} \left\{      \frac{\chi_{\rm o} + i \chi_{\rm e}}{\mu_0}\,
 \textbf{B}_0\, i e^{-i \omega_0 t}   \right\} \right)
 \Cdot\Nabla\right]{\bf B}_{\rm app}\right\rangle V
\label{eq:Force}
\end{equation}
and
\begin{eqnarray}
 \textbf{N} & = & \Big\langle \textbf{M} \times \textbf{B}_{\rm app}
 +\textbf{r} \times \Big([\textbf{M} \cdot \Nabla ] \textbf{B}_{\rm app}\nonumber\\
 & + & \left[ {\rm Re} \left\{      \frac{\chi_{\rm o} + i \chi_{\rm e}}{\mu_0}\,\textbf{B}_0\, i e^{-i \omega_0 t}   \right\} \cdot 
 \Nabla \right] \textbf{B}_{\rm app}\Big) \Big\rangle V
\label{eq:Torque}
\end{eqnarray}
where  {\bf B}$_{\rm app}$  is the  field produced  by the  coils, and
$\textbf{r}$ is the  position vector that has the  test mass center as
origin. The  term ${\rm Re}  \left\{ \frac{\chi_{\rm o} +  i \chi_{\rm
e}}{\mu_0}\,\textbf{B}_0\,  i  e^{-i  \omega_0  t}  \right\}$  is  the
induced magnetization  by the externally applied field.  Note that the
forces and  torques depend on  {\bf M}, $\chi_{\rm o}$  and $\chi_{\rm
e}$.

Considering
Eqs.~(\ref{eq.appfield}),~(\ref{eq.appfield2})~and~(\ref{eq:Force}),
the $x$-component of the force acting on the test mass is

\begin{eqnarray}
\label{eq.forces}
 F_x  & = &  \frac{\chi_{\rm o} V}{2\mu_0}\,\left\langle
 {\bf B}_0\Cdot\Nabla B_{0,x}\right\rangle \nonumber\\
&+& \left\langle{\bf M}\Cdot\Nabla B_{0,x}\right\rangle 
 V\,\sin\omega_0 t \nonumber \\
  & -&   \frac{\chi_{\rm o} V}{2\mu_0}\,\left\langle
 {\bf B}_0\Cdot\Nabla B_{0,x}\right\rangle\,\cos\,2\omega_0 t  \nonumber \\
 & -&   \frac{\chi_{\rm e} V}{2\mu_0}\,\left\langle
 {\bf B}_0\Cdot\Nabla B_{0,x}\right\rangle\,\cos(2\omega_0 t - \pi/2) 
\end{eqnarray}

\noindent where we have  used that $\sin^2\omega_0 t=(1-\cos 2\omega_0
t)/2$ and  the $\pi/2$ rad phase  due to the complex  component of the
susceptibility  has been  added as  an argument  in  the corresponding
$\cos$  term.   As  can  be   seen  from  this  equation,  the  linear
acceleration of  the test masses  along the $x$-axis has  two separate
frequencies, one at $\omega_0$ and  the other at $2\omega_0$, and also
a  DC  component.  The  $2\omega_0$  component  presents  an  in-phase
component proportional  to $\chi_{\rm  o}$ and a  quadrature component
proportional   to  $\chi_{\rm   e}$.   Particularly,   the  $\omega_0$
component can be more explicitly written as:

\begin{equation}
 \left\langle{\bf M}\Cdot\Nabla B_{0,x}\right\rangle = 
 \left\langle M_x\frac{\partial B_{0,x}}{\partial x} + 
              M_y\frac{\partial B_{0,x}}{\partial y} + 
              M_z\frac{\partial B_{0,x}}{\partial z}\right\rangle
\end{equation}

\noindent where $M_x$, $M_y$, $M_z$  are the components of the density
of the remanent  magnetic moment.  If the test  mass is homogeneous we
have the simplified expression

\begin{equation}
 \left\langle{\bf M}\Cdot\Nabla B_{0,x}\right\rangle = 
 \left\langle M_x\frac{\partial B_{0,x}}{\partial x} \right\rangle
\end{equation}

\noindent  since the $y\/$  and $z\/$  components of  $\Nabla B_{0,x}$
average to zero  due to symmetry of the field of  the coil. This leads
to a  force component along the  $x$-axis that only  depends on $M_x$,
$\chi_{\rm o}$ and $\chi_{\rm e}$.

On  the other hand,  the torque  acting on  the test  mass also  has a
similar behavior:
\begin{eqnarray}
 \mathbf{N}  &=&  \left\langle\mathbf{M}\times\mathbf{B}_0 +
 \mathbf{r}\times\left[\left(\mathbf{M}\Cdot\Nabla\right)\mathbf{B}_0\right] 
 \right\rangle V\,\sin\omega_0 t\label{eq.9} \\ 
 &+&  \left\langle  \mathbf{r}\times
     \frac{\chi_{\rm o}}{\mu_0}\,
 \left[\mathbf{B_0} \cdot \Nabla] \mathbf{B_0} \right]  
 \right\rangle  V\,\sin^2\omega_0 t \nonumber \\
 &-&  \left\langle  \mathbf{r}\times
     \frac{\chi_{\rm e}}{\mu_0}\,
 \left[\mathbf{B_0} \cdot \Nabla] \mathbf{B_0} \right]  
 \right\rangle  V\,\sin\omega_0 t \, \cos\omega_0 t  \nonumber
\end{eqnarray}
In this  case, it must be noted  that, because of the  symmetry of the
applied magnetic  field, the terms multiplying  $\sin^2\omega_0 t$ and
$\sin \omega_0 t \, \cos  \omega_0 t$ in Eq.~(\ref{eq.9}) vanish.  The
two rotation excursions detected by the interferometer using wavefront
sensing  are  the rotations  about  the  $y$-axis  and $z$-axis.   The
magnitude of the rotation about the $x$-axis is smaller, and cannot be
detected by the interferometer because the axis of rotation is aligned
with  the laser  beam.  Taking this  into  account,  the two  relevant
torques for the experiment are:

\begin{eqnarray}
\label{eq.torq1}
N_{y} & = & \Big\langle M_z B_{0,x} - M_x B_{0,z}\nonumber\\
& + & z \, ( {\bf M} \Cdot \Nabla B_{0,x} )\nonumber\\
& - & x \, ( {\bf M} \Cdot \Nabla B_{0,z} )   
\Big \rangle V \,\sin\omega_0 t \\
\label{eq.torq2}
N_{z} & = & \Big\langle M_x B_{0,y} - M_y B_{0,x}  \nonumber\\
& + & x \, ( {\bf M} \Cdot \Nabla B_{0,y} )\nonumber\\
& - & y \, ( {\bf M} \Cdot \Nabla B_{0,x} )   
\Big \rangle V \,\sin\omega_0 t 
\end{eqnarray}

These equations can be further simplified in the case of a homogeneous
test mass.   In this case, due  to the axial symmetry  of the magnetic
field, the terms  $\langle B_{0,z}\rangle$ in Eq.~(\ref{eq.torq1}) and
$\langle  B_{0,y}\rangle$ in  Eq.~(\ref{eq.torq2})  vanish.  Moreover,
for the same reason the terms
$$\left\langle z \, \frac{\partial B_{0,x}}{\partial x} \right\rangle,~~~~
  \left\langle z \, \frac{\partial B_{0,x}}{\partial y} \right\rangle,$$ 
and
$$\left\langle x \, \frac{\partial B_{0,z}}{\partial x} \right\rangle,~~~~
  \left\langle x \, \frac{\partial B_{0,z}}{\partial y} \right\rangle$$ 
in Eq.~(\ref{eq.torq1}) also vanish, as do the terms 
$$\left\langle x \, \frac{\partial B_{0,y}}{\partial x} \right\rangle,~~~~
  \left\langle x \, \frac{\partial B_{0,y}}{\partial z} \right\rangle,$$ 
$$\left\langle y \, \frac{\partial B_{0,x}}{\partial x} \right\rangle,~~~~ 
  \left\langle y \, \frac{\partial B_{0,x}}{\partial z} \right\rangle$$ 
in  Eq.~(\ref{eq.torq2}).  In  these terms  $x$, $y$  and $z$  are the
three components of $\textbf{r}$. Hence, the torque about the $y$-axis
only depends on  $M_z$ and the torque about  the $z$-axis only depends
on $M_y$:

\begin{eqnarray}
N_{y}  =  M_z \, \Big\langle  B_{0,x} +  z \,  
\frac{\partial B_{0,x} }{\partial z}  - x \,  
\frac{\partial B_{0,z} }{\partial z}  \Big \rangle V \,\sin\omega_0 t 
\label{eq.torques1}\\
N_{z}  =  M_y \, \Big\langle - B_{0,x} +  x \,  
\frac{\partial B_{0,y} }{\partial y}  - y \,  
\frac{\partial B_{0,x} }{\partial y}  \Big \rangle V \,\sin\omega_0 t 
\label{eq.torques2}
\end{eqnarray}

Finally, we  can cast Eqs.~(\ref{eq.forces}),  (\ref{eq.torques1}) and
(\ref{eq.torques2}) in the form:

\begin{eqnarray}
\label{eq.structure}
 F_x &  = &\chi_{\rm o} ~ f_{x_{\rm DC}} + M_x ~ f_{x_{1\omega_0}} + 
           \chi_{\rm o} ~ f_{x_{2\omega}} + \chi_{\rm e} ~ f^{''}_{x_{2\omega_0}} \nonumber\\
 N_y &  = & M_z ~ n_{y_{1\omega_0}} \\ 
 N_z &  = & M_y ~ n_{z_{1\omega_0}} \nonumber 
\end{eqnarray}

\noindent   where   $f_{x_{\rm   DC}}$   is   a   constant   function,
$f_{x_{1\omega_0}}$,   $n_{y_{1\omega_0}}$   and   $n_{z_{1\omega_0}}$
oscillate     at     $\omega_0$     and    $f_{x_{2\omega_0}}$     and
$f^{''}_{x_{2\omega_0}}$ oscillate at $2\omega_0$.


\section{Dynamic model}
\label{chap.3}

The LTP instrument  will react to the injection  of the aforementioned
forces and torques inflicted upon the test masses. This will result in
specific kinematic  excursions in  both test masses.   These kinematic
excursions will depend  on the instrument dynamics and  will be sensed
by the onboard  interferometer.  The LTP is a  very complex instrument
and     its    modeling    has     been    presented     in    several
references~\cite{bib:mock,bib:MIMO,bib:control}. It  can be modeled by
splitting it in four main subsystems which are:

\begin{enumerate}
 \item The dynamical  model ({\bf D}) represents the  evolution of the
   kinematic excursions of  the two test masses placed  inside the LTP
   and the kinematics of the spacecraft. This model takes into account
   the coupling  of the  motion of  each of the  test masses  with the
   motion  of the  spacecraft  and  outputs the  evolution  of the  15
   degrees of freedom of the instrument (6 for each of the test masses
   and 3 more for the spacecraft).

 \item  The sensing  mechanisms ({\bf  S})  onboard LPF  are the  star
   tracker, the inertial sensors,  and the interferometer.  Of special
   interest is the interferometer, which measures the distance between
   test  mass  1   and  the  spacecraft  and  between   the  two  test
   masses~\cite{bib:interferometer}.

 \item The  controller blocks ({\bf  C}) are in charge  of calculating
   the  appropriate commands  to  correct the  positions  of the  test
   masses and the  attitude of the spacecraft. In  science mode, there
   are two  main control loops  applied by the instrument.   The first
   one ---  the drag free loop  --- takes the  absolute measurement of
   the distance between test mass 1 and the spacecraft as a reference.
   It then calculates which forces should be applied to the spacecraft
   in order  to counteract all  disturbances and recreate a  drag free
   environment for test mass 1.  The second loop --- the low frequency
   loop --- takes as  a reference the differential measurement between
   both test  masses and  acts on  the second test  mass to  avoid its
   collision   with  the   spacecraft   walls~\cite{bib:control}.  The
   controllers have  been designed to deliver  very sensitive readings
   of the differential  motion of both test masses  between 1\,mHz and
   30\,mHz,   the   measurement   bandwidth   of   the   LPF   mission
   \cite{bib:trento}.  These two  control loops are implemented inside
   the onboard computer of the LPF.

\item  The actuators  ({\bf A})  are the  physical systems  that apply
  these commands  to the test masses  and to the  spacecraft.  The two
  actuator   mechanisms   existing    in   LPF   are   the   satellite
  micropropulsion system,  which is  composed by 12  micro-newton FEEP
  thrusters (Field  Emission Electric Propulsion),  and the capacitive
  actuators which  consist of  a set of  electrodes that  surround the
  test masses and exert controlled forces on them.

\end{enumerate}

\begin{figure}[t]
 \centering
 \includegraphics[width = \columnwidth]{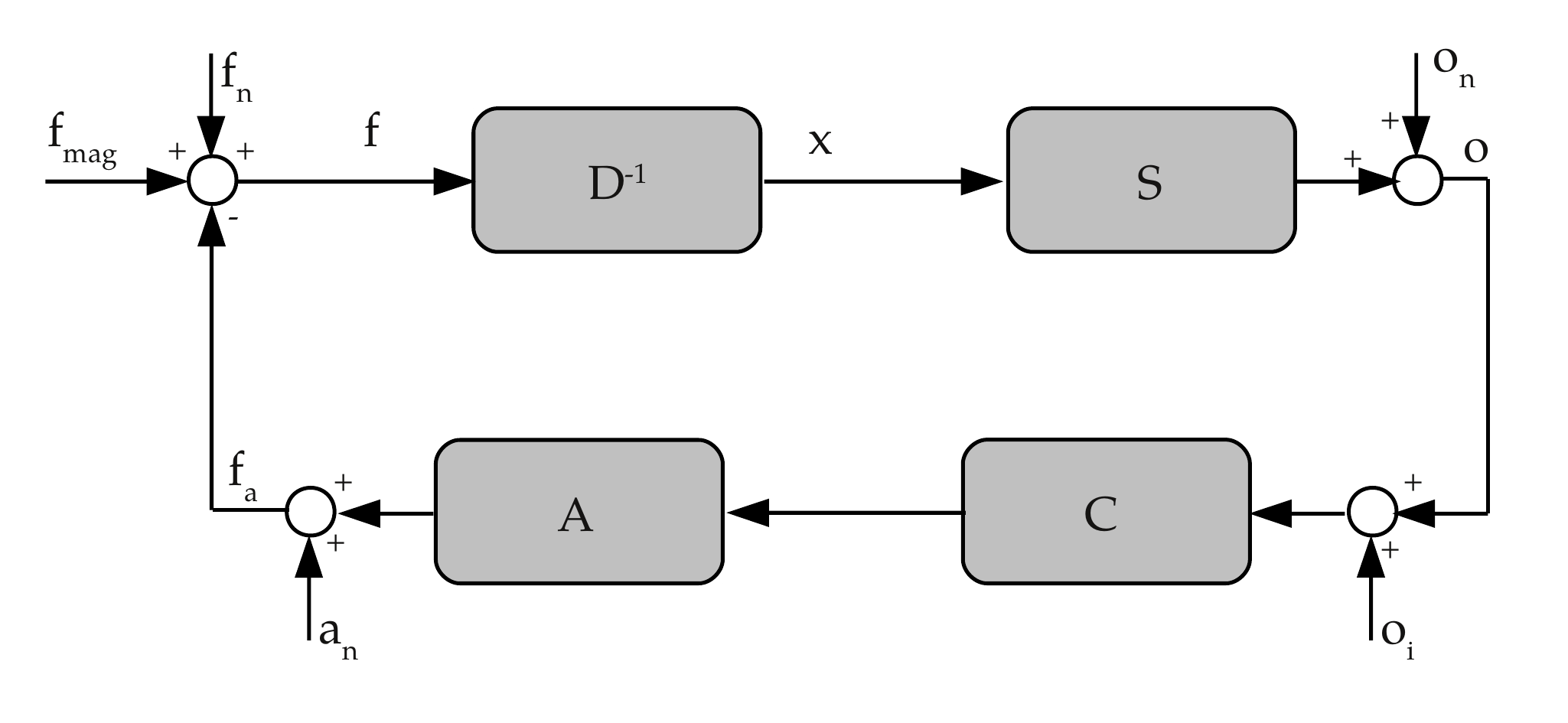}
 \caption{Control  system  architecture  of  LISA  Pathfinder.   ${\bf
   D}$~stands  for  the  dynamical  matrix, ${\bf  S}$~represents  the
   sensing matrix of the  interferometer, i.e.  the matrix translating
   the position of  the test mass, ${\bf x}$,  into the interferometer
   readout,  ${\bf  o}$ (${\bf  o}_{\rm  n}$  stands  for the  readout
   noise).   ${\bf A}$~represents  the physics  of the  FEEP  and the
   electrostatic actuators,  and finally ${\bf C}$,  is the controller
   matrix, implementing the drag free and low-frequency control loops.
   ${\bf  o}_{\rm i}$  represents the  displacement  guidance signals.
   ${\bf a}_{\rm n}$ are the actuators noise and ${\bf f}_{\rm a}$ are
   the output forces and torques of the actuators. ${\bf f}_{\rm mag}$
   are the magnetic forces and  torques induced by the coils and ${\bf
   f}_{\rm n}$ are the  environment force and torque noises disturbing
   the spacecraft.}
 \label{fig.control}
\end{figure}

This subsystem division is schematically shown in the block diagram of
Fig.~\ref{fig.control}. For  more detailed information  of the system,
the reader is referred to Refs.~\cite{bib:mock,bib:control,bib:MIMO}.

In the magnetic  experiment the input signals are  the magnetic forces
and torques (${\bf f}_{\rm mag}$), and the outputs are the readings of
the interferometer  (${\bf o}$).   Therefore, using the  block diagram
scheme  shown in Fig.~(\ref{fig.control}),  we calculate  the transfer
function, which results in:
\begin{equation}
{\bf o} = {\bf H} \cdot {\bf f_{mag}}
\end{equation}
where 
\begin{equation}
{\bf H} = \frac{{\bf D}^{-1} \cdot 
 {\bf S}}{1 + {\bf D}^{-1} \cdot {\bf S} \cdot {\bf A} \cdot {\bf C}}
\label{eq.transferfunction}
\end{equation}
This transfer  function depends on all the  above described subsystems
and  represents  the  dynamical  response  of the  instrument  to  the
specific injected signals.


\section{Estimation model}
\label{chap.4}

The estimation of the magnetic characteristics is performed processing
the interferometer readings. To do  so, we use the displacement of the
differential channel  ($o_{x_{12}}$), the rotation  about the $y$-axis
($o_{\eta_1}$) and the rotation about the $z$-axis ($o_{\phi_1}$).  If
cross talks  are disregarded, the reading of  the displacement channel
stems  only by  the  effect of  the  magnetic force  acting along  the
$x$-axis, $F_x$.   Analogously, something  similar occurs for  the two
torques in each of their respective axis. Thus, we can write:

\begin{eqnarray}
 \label{eq.crosstalkfree}
 o_{x12} & = & M_x~d_{x_{1\omega_0}} + 
             \chi_{\rm o}~d_{x_{2\omega_0}} + 
             \chi_{\rm e}~d^{''}_{x_{2\omega_0}} \nonumber  \\
 o_{\eta_1} & = & M_z~r_{y_{1\omega_0}} \\
 o_{\phi_1} & = & M_y~r_{z_{1\omega_0}} \nonumber
\end{eqnarray}

\noindent    where   $d_{x_{1\omega_0}}$,    $d_{x_{2\omega_0}}$   and
$d^{''}_{x_{2\omega_0}}$ are the respective transformations from force
to     displacement     of     the    signals     $f_{x_{1\omega_0}}$,
$f_{x_{2\omega_0}}$        and       $f^{''}_{x_{2\omega_0}}$       in
Eq.~(\ref{eq.structure}), and  analogously for $r_{y_{1\omega_0}}$ and
$r_{z_{1\omega_0}}$   for   the   case  of   $n_{y_{1\omega_0}}$   and
$n_{z_{1\omega_0}}$.  Nevertheless, because  of the high complexity of
the  LTP instrument,  this model  is not  sufficiently  realistic.  In
particular, it  turns out that the cross-talks  cause important biases
in  the  parameter  estimates.  This  is because  the  effect  of  the
$x$-force in  the rotation readings and  the effect of  the torques in
the $x$-axis readings  are not negligible.  As a  consequence, we used
the full three-dimensional model of the experiment:

\begin{eqnarray}
\label{eq.estimationmodel}
 \left(
\begin{array}{c}
 o_{x_{12}} \\
 o_{\eta_{1}} \\
 o_{\phi_{1}} \\
\end{array}
\right) &\,& = \left(
\begin{array}{ccc}
 H_{F_x \rightarrow x_{12}} & H_{N_y \rightarrow x_{12}} & H_{N_z \rightarrow x_{12}} \\
 H_{F_x \rightarrow \eta_{1}} & H_{N_y \rightarrow \eta_{1}} & H_{N_z \rightarrow \eta_{1}} \\
 H_{F_x \rightarrow \phi_{1}} & H_{N_y \rightarrow \phi_{1}} & H_{N_z \rightarrow \phi_{1}} \\
\end{array}
\right) \cdot \nonumber \\
&\,& \left(
\begin{array}{ccccc}
 M_x & \chi_{\rm o} & \chi_{\rm e} & 0 & 0 \\
 0 & 0 & 0 & M_z & 0 \\
 0 & 0 & 0 & 0 & M_y \\
\end{array}
\right) \left(
\begin{array}{c}
 f_{x_{1\omega_0}} \\
 f_{x_{2\omega_0}} \\
 f^{''}_{x_{2\omega_0}} \\
 n_{y_{1\omega_0}} \\
 n_{z_{1\omega_0}}
\end{array}
\right)
\end{eqnarray}

\noindent where the $3 \times  3$ matrix {\bf H} is the transformation
matrix from force/torque  to displacement/rotation that represents the
closed     loop    dynamics    of     the    instrument     ---    see
Eq.~(\ref{eq.transferfunction}). This matrix is not diagonal, as it is
assumed  in the  model  in  which the  cross-talks  are neglected  ---
namely, Eq.~(\ref{eq.crosstalkfree}). For  instance, the effect of the
torque  about  the  $y$-axis  and  the $z$-axis  on  the  $o_{x_{12}}$
displacement channel is relevant, and thus non-zero transfer functions
$H_{N_y \rightarrow x_{12}}$ and  $H_{N_z \rightarrow x_{12}}$ need to
be considered.  Hence, to estimate $M_x$, $M_y$, $M_z$, $\chi_{\rm o}$
and $\chi_{\rm  e}$, these transfer  functions have to be  known. This
model is  still a  simplification, because we  do not include  all the
degrees of  freedom, but it is  certainly more realistic  than that of
Eq.~(\ref{eq.crosstalkfree}), which is strictly one-dimensional.

\subsection{Estimation procedure and bias correction}

The   estimation    procedure   has   been    already   described   in
Ref.~\cite{bib:myPaper3}.   However,  in  this  paper  we  present  an
important  modification  to  correct  for  the  biases  introduced  by
cross-talks.  The  full three-dimensional  estimation  model given  by
Eq.~(\ref{eq.estimationmodel}) may be regrouped as:

\begin{eqnarray}
 o_{x12} & = & (M_x + \alpha_{12} M_z + \alpha_{13} M_y)~d_{x_{1\omega_0}} + \nonumber \\
        & + &\chi_{\rm o}~d_{x_{2\omega}} + \chi_{\rm e}~d^{''}_{x_{2\omega_0}}\nonumber \\ 
        \label{chap4:eq3D}
 o_\eta & = & (\alpha_{21}M_x + M_z + \alpha_{23} M_y)~r_{y_{1\omega_0}}  \\  
 o_\phi & = & (\alpha_{31} M_x + \alpha_{32} M_z + M_y)~r_{z_{1\omega_0}} \nonumber
\end{eqnarray}

\noindent where $\alpha$ are the cross-talks of the system (the matrix
elements of  {\bf H}  evaluated at the  excitation frequency).   If we
introduce primed quantities  $M_x^{'}$, $M_y^{'}$, and $M_z^{'}$, then
Eq.~(\ref{chap4:eq3D}) can be written as:

\begin{eqnarray}
 o_{x12} & = & M_x^{'}~d_{x_{1\omega_0}} + \chi_{\rm o}~d_{x_{2\omega_0}} 
 + \chi_{\rm e}~d^{''}_{x_{2\omega_0}} \nonumber \\
 o_\eta & = & M_z^{'}~r_{y_{1\omega_0}} \\
 o_\phi & = & M_y^{'}~r_{z_{1\omega_0}} \nonumber
\end{eqnarray}

\noindent   and   we   estimate   the   values   of   $\hat{M}_x^{'}$,
$\hat{M}_y^{'}$,    $\hat{M}_z^{'}$,    $\hat{\chi}_{\rm    o}$    and
$\hat{\chi}_{\rm  e}$  applying standard  single  output least  square
techniques  \cite{bib:myPaper3,  bib:leastSquares}.   These values  of
$\hat{M}_x^{'}$, $\hat{M}_y^{'}$  and $\hat{M}_z^{'}$ are  biased, and
do not correspond to the true magnetic moment components, $M_x$, $M_y$
and $M_z$. Nevertheless, these biases can be corrected because we know
the relation between them:

\begin{equation}
\left(
\begin{array}{c}
\hat{M}_x \\
\hat{M}_z \\
\hat{M}_y \\ 
\end{array}
\right)
=
\left(
\begin{array}{ccc}
1 & \alpha_{12} & \alpha_{13} \\
\alpha_{21} & 1 & \alpha_{23} \\
\alpha_{31} & \alpha_{32} & 1 \\
\end{array}
\right)^{-1}
\left(
\begin{array}{c}
\hat{M}_x^{'} \\
\hat{M}_z^{'} \\
\hat{M}_y^{'} \\ 
\end{array}
\right)
\end{equation}

Note  that Eq.~(\ref{eq.estimationmodel}) provides  the values  of the
elements    of    this    matrix,    and   that    the    matrix    is
invertible. Additionally, it is worth emphasizing that we only correct
the components of the magnetic  moment and no correction is considered
for the  susceptibility ($\chi_{\rm o}$ and $\chi_{\rm  e}$).  This is
because the magnetic susceptibility is not affected by any cross-talk.
It turns out that the previously outlined procedure corrects biases of
around  1\% in  each of  the magnetic  parameters, which  are sizable.
Finally, we also mention that  during the lifetime of the mission some
of the  telemetry channels  may fail.  Thus,  it is important  to know
beforehand that  single channel estimation is still  possible and that
it introduces biases of $\sim 1\%$. On the other hand, by direct usage
of  the LTPDA  toolbox, and  in order  to avoid  additional estimation
biases induced  by the  low frequency behavior  of the  instrument, we
whiten the data and we  eliminate its transients. These techniques are
expected  to be  used  in other  experiments  of the  mission and  are
described   in   detail  elsewhere~\cite{bib:mock,   bib:DataAnalysis,
bib:LTPDA,bib:miquel}.


\section{An analysis of the uncertainties}
\label{chap.5}

The Experiment Master Plan of  the mission is aimed at determining the
physical parameters of the  instrument, characterizing in this way the
matrix elements of {\bf H}. These transfer functions depend on several
physical parameters.   Amongst them we mention the  stiffnesses of the
test masses ($\omega_1$  and $\omega_2$, where the subindexes  1 and 2
refer,  respectively,  to test  mass  1  and  2), the  actuator  gains
(namely, the  gain of the FEEP  actuator, $G_{\rm FEEP}$,  and that of
the  capacitive  actuator,   $G_{\rm  CA}$),  and  the  interferometer
cross-coupling  ($\delta_{12}$).   In  the  end,  this  results  in  a
complete characterization  of the main four blocks  of the instrument.
Detailed information on the Experiment Master Plan and on the accuracy
of       the        estimates       can       be        found       in
Refs.~\cite{bib:mock,bib:DataAnalysis,bib:Nofrarias}.

Nevertheless, for  the calculations presented here it  is important to
realize  that  some of  the  parameters of  the  model  may be  poorly
determined or have sizable  uncertainties.  Therefore, in our analysis
we  introduce  uncertainties  in  each  of  the  most  relevant
parameters of the mission. These uncertainties are represented as $b$,
and  the  subscript  ``NOM''  stands  for the  nominal  value  of  the
parameter:

\begin{eqnarray}
\omega_1 &=& \omega_{1_{\rm NOM}} ( 1 \pm b_{\omega_1})\nonumber\\
\omega_2 &=& \omega_{2_{\rm NOM}} ( 1 \pm b_{\omega_2})\nonumber\\
\delta_{12} &=& \delta_{{12}_{\rm NOM}} ( 1 \pm b_{\delta_{12}})\nonumber\\
G_{\rm FEEP} &=& G_{{\rm FEEP}_{\rm NOM}} ( 1 \pm b_{G_{\rm FEEP}})\nonumber\\
G_{\rm CA} &=& G_{{\rm CA}_{\rm NOM}} ( 1 \pm b_{G_{\rm CA}})\nonumber
\end{eqnarray} 

Clearly, the effects  of these uncertainties on the  estimation of the
magnetic parameters need to be assessed.  To this end, for each of the
nine transfer functions  of ${\bf H}$, we have  computed the effect of
the uncertainties  on each of the  parameters of the  system.  We have
done this analysis for values of $b$ ranging from $-0.2$ to $0.2$, and
we have  studied their effect on the  modulus and on the  phase of the
transfer  functions.   We  have  found  that the  uncertainty  on  the
capacitive actuator gain ($b_{G_{\rm CA}}$) is the only one that has a
relevant impact, whilst the uncertainties on the other parameters have
a negligible effect.

\subsection{The gain of the capacitive actuator}
\label{sec:sensitivity}

\begin{figure}[!t]
 \centering
 \includegraphics[width = \columnwidth]{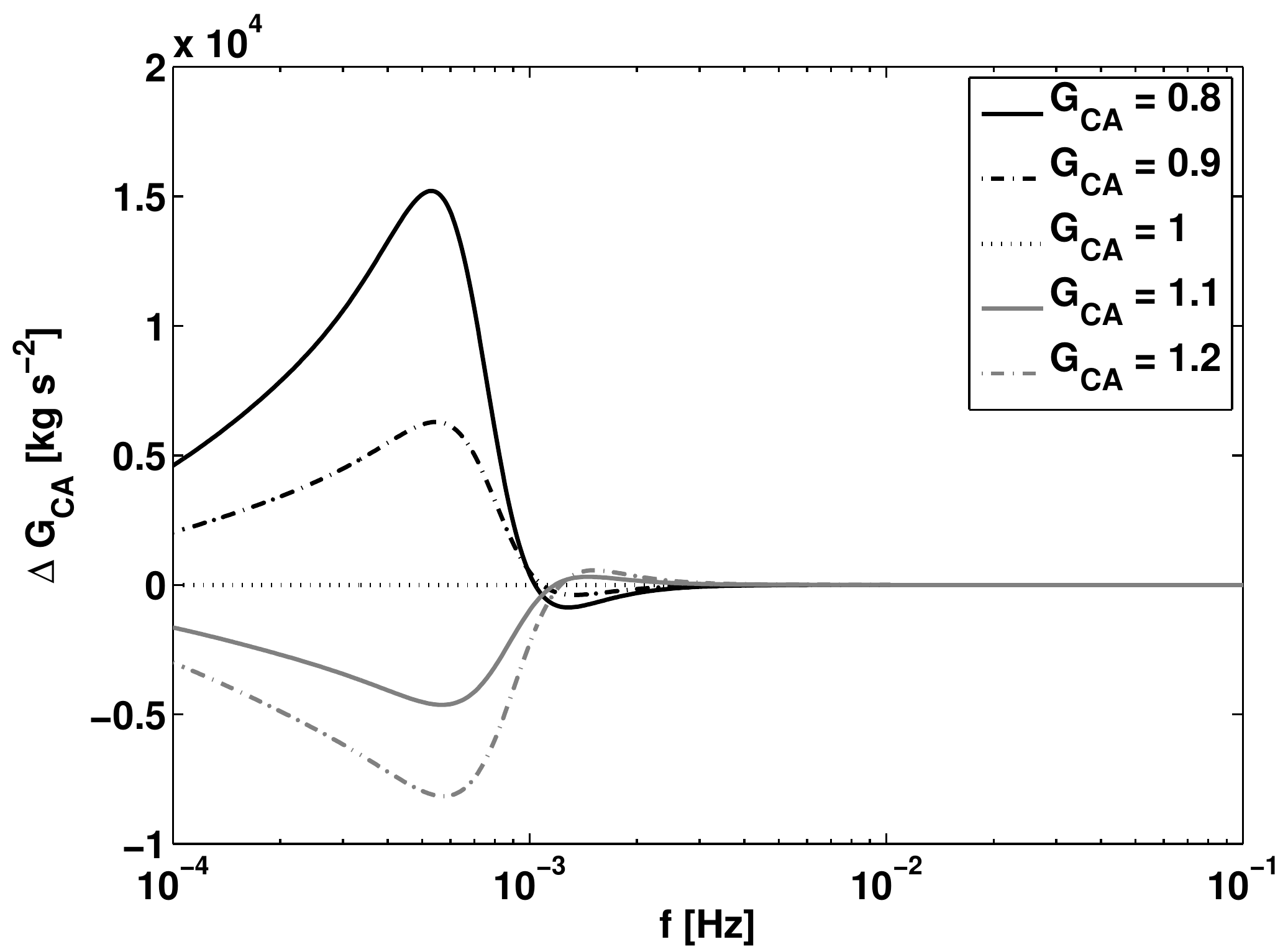}
 \includegraphics[width = \columnwidth]{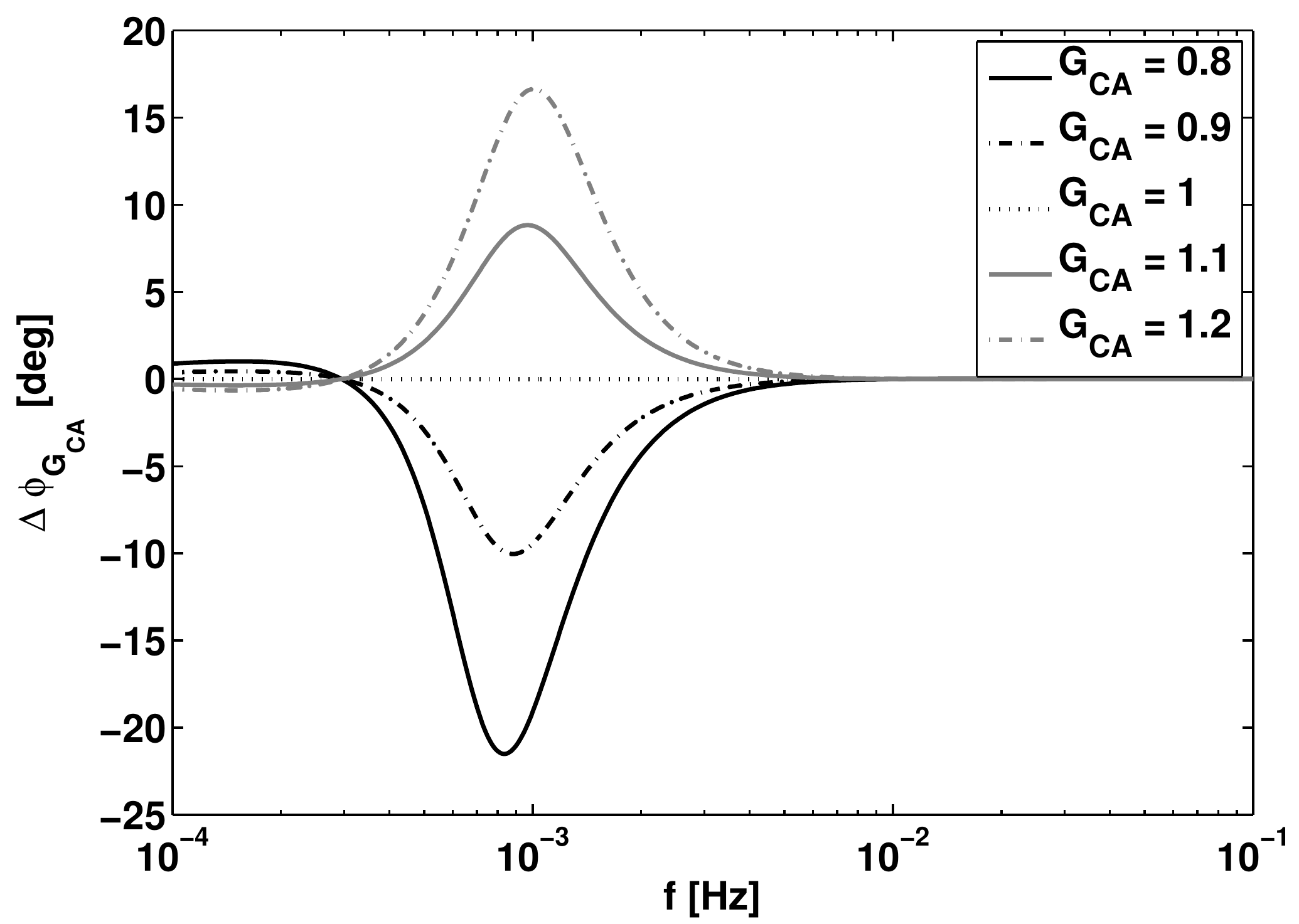}
 \caption{Top  panel: error of  the modulus  of the  transfer function
   $H_{F_x  \rightarrow   x_{12}}^b$  with  respect   to  its  nominal
   behavior.  This  frequency-dependent relative error  is plotted for
   different  capacitive  actuator  gain  uncertainties  ranging  from
   $-0.2$  to 0.2.   Bottom panel:  phase differences  in  the $H_{F_x
   \rightarrow    x_{12}}^b$   transfer    function    for   different
   uncertainties of  the gain of  the capacitive actuator.   The phase
   differences  are  also   calculated  for  different  relative  gain
   uncertainties ranging from $-0.2$ to 0.2.}
 \label{fig:sensitivity}
\end{figure}

In  this section  we  analyze the  effect  of the  uncertainty of  the
capacitive  actuator gain  ($b_{G_{\rm  CA}}$).  To  this  end, for  a
specific value of  $b_{G_{\rm CA}}$, we compute the  absolute error of
the  modulus  (${\bf  H_{\rm  e}}^{\rm  b}$) of  the  system  transfer
functions and  the phase  differences (${\bf H_\psi}^{\rm  b}$) across
the measurement bandwidth (1\,mHz to 30\,mHz):

\begin{eqnarray}
 {\bf H_{\rm e}}^{\rm b} & = & |{\bf H }^{\rm b}| - |{\bf H }| \\
 {\bf H_\psi}^{\rm b} & = & \measuredangle {\bf H }^{\rm b} - 
                         \measuredangle {\bf H }
\end{eqnarray}

\noindent where  $\measuredangle$ stands for the  matrix operator that
calculates the phase  of each of the elements of  the matrix. In these
expressions,  the  superscript   ``b''  indicates  that  the  specific
transfer function  has been  calculated with a  non-zero value  of the
parameter uncertainty.  On the contrary, functions without superscript
have  been  calculated with  the  nominal  values  of all  the  system
parameters.  Therefore,  ${\bf  H}_{\rm  e}^{\rm  b}$  calculates  the
absolute  error of  the modulus  of each  of the  nine  functions with
respect   to   its  nominal   value   for   one   specific  value   of
uncertainty~($b$), and ${\bf H_\psi}^{\rm  b}$ gives account for their
phase differences.  These two  matrices give a quantitative assessment
of the error  of the model due to the  uncertainties across the entire
measurement  bandwidth.  Moreover, the  most relevant  contribution in
the error of the model will be  due to the error in the diagonal terms
of the  matrix.  Therefore, we  analyze mainly the effects  on $H_{F_x
\rightarrow  x_{12}}$,  $H_{N_y  \rightarrow  \eta_{1}}$  and  $H_{N_z
\rightarrow \phi_{1}}$.

Fig.~\ref{fig:sensitivity}  displays the  results of  this sensitivity
analysis for  $H_{F_x \rightarrow x_{12}}$,  i.e the first  element of
matrices  ${\bf H_{\rm  e}}^{\rm b}$  and ${\bf  H_\psi}^{\rm  b}$ for
several values  of the uncertainty  in the capacitive  actuator gains,
ranging  from  $-0.2$ to  0.2.   The behavior  as  a  function of  the
frequency of the other two  elements of the diagonal are very similar.
In the top panel  of this figure it can be seen  that the error of the
modulus is especially relevant  below 1\,mHz, where the differences in
amplitude increase up  to 48\% for 0.6\,mHz, when  the capacitive gain
is  0.8 (instead  of 1).   The changes  in modulus  are  also relevant
between  1\,mHz and  ~7\,mHz.  In  the  bottom panel,  we examine  the
differences in  the phase  of the same  transfer functions. It  can be
seen  that there  exist phase  shifts of  15$^\circ$ for  a capacitive
actuator gain of 1.2 at a frequency of ~1\,mHz. These phase shifts are
relevant  between  0.4\,mHz  and  4\,mHz.   Such  differences  produce
important  biases  in  the   estimates  of  the  magnetic  parameters.
Moreover, the  effect depends on the excitation  frequency.  Thus, the
choice  of the right  excitation frequency  ($\omega_0$) is  a crucial
aspect  in  the  experiment  design.   We postpone  this  analysis  to
Sect.~\ref{chap.6},  where   we  will  study  which   is  the  optimal
excitation frequency.  Finally, we  also mention that similar analyses
for the rest of the  uncertainties on the nominal parameters have been
performed, but are not shown here for the sake of conciseness.


\section{The optimal frequency}
\label{chap.6}

Finding the  optimal frequency of the sinusoidal  currents injected in
the coils to obtain the magnetic  parameters is a crucial issue of the
experiment.   Actually,  as  it  will  be  shown  below,  the  optimal
frequency can be obtained from a trade-off between the frequency range
where the  instrument presents a maximum of  the signal-to-noise ratio
(SNR) and the  frequency range where the instrument  is less sensitive
to  the  uncertainties  of   the  capacitive  actuator  gain  ---  see
Sect~\ref{sec:sensitivity}.

\begin{figure}[!t]
 \centering
 \includegraphics[width= \columnwidth]{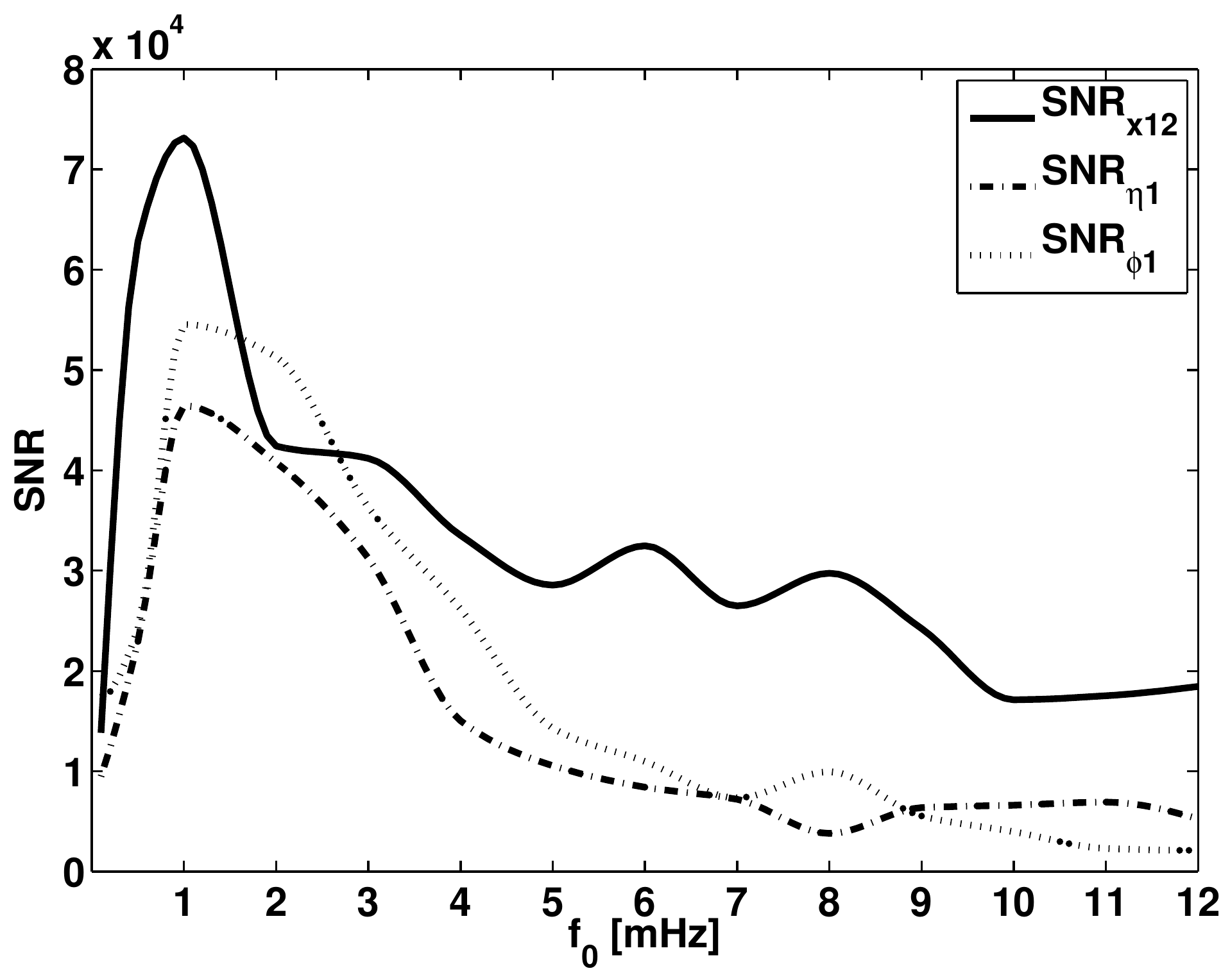}
 \caption{Signal-to-noise ratio as a function of frequency for each of
   the relevant  signals of the magnetic  experiment, the differential
   channel,  $x_{12}$  ---  solid  line  --- the  rotation  about  the
   $y$-axis,  $\eta_1$ ---  dashed-dotted  line ---  and the  rotation
   about the $z$-axis, $\phi_1$ --- dotted line.}
 \label{fig:SNR}
\end{figure}

\begin{figure}[!t]
 \centering
 \includegraphics[width = \columnwidth]{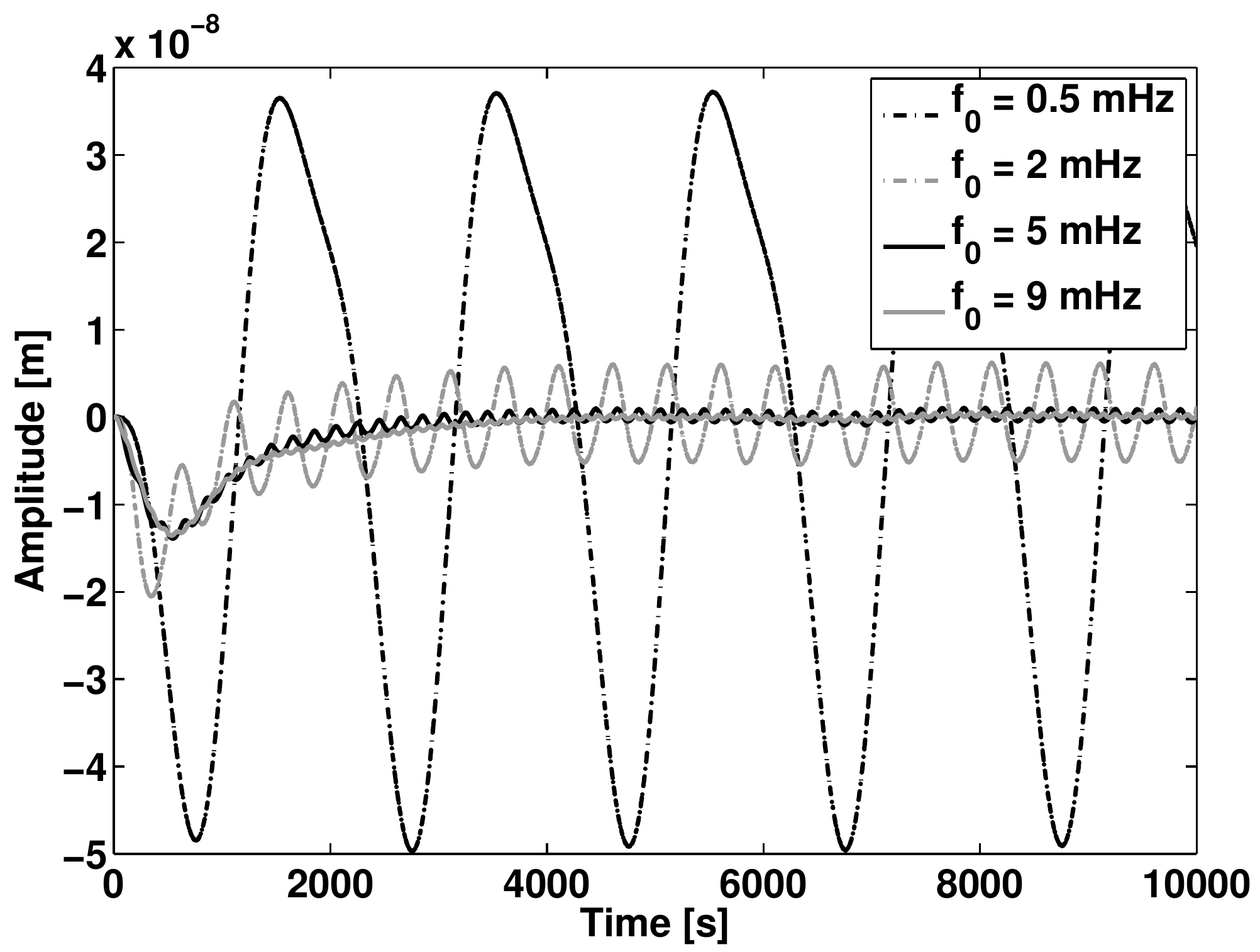}
 \includegraphics[width = \columnwidth]{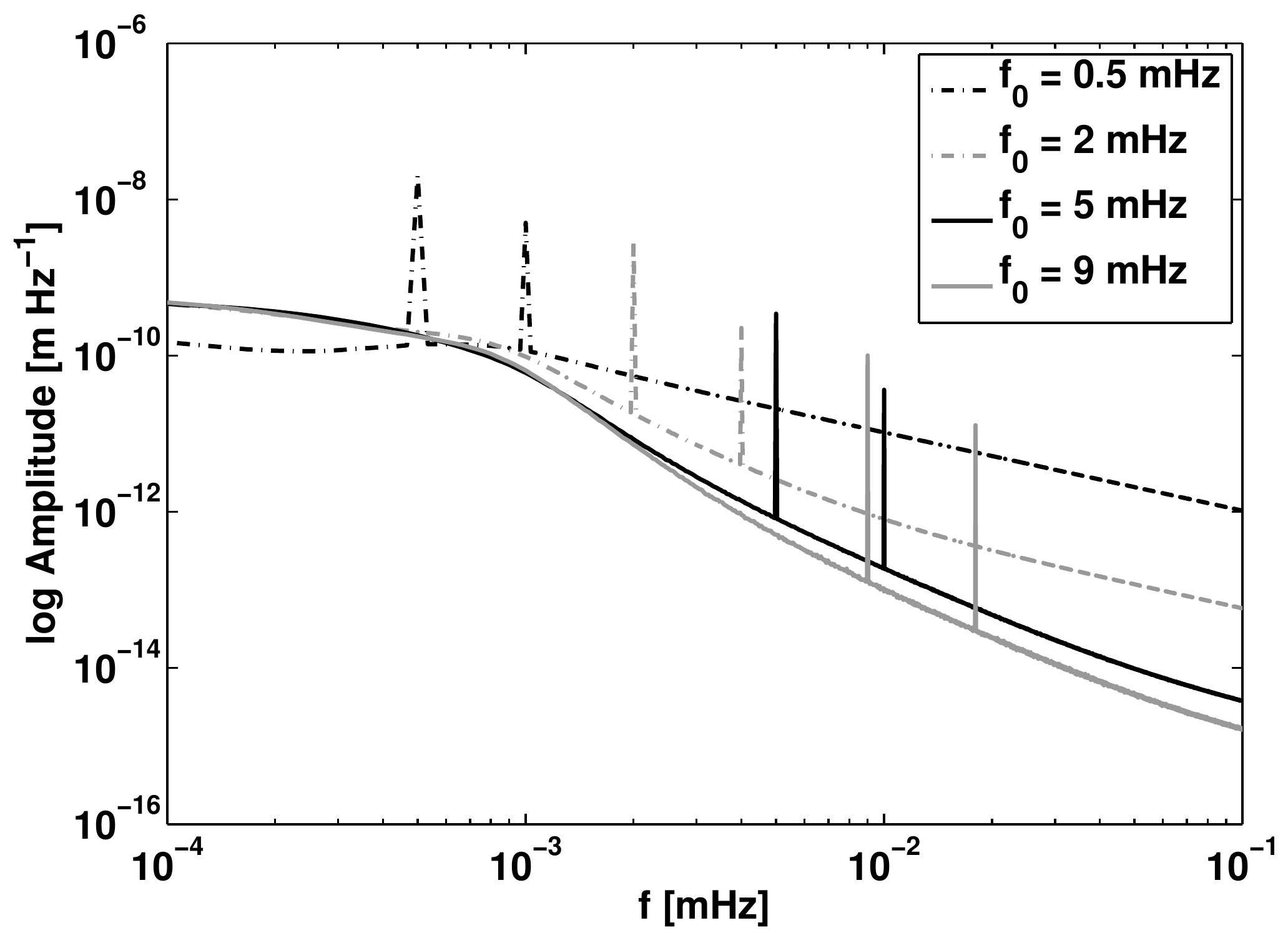}
 \caption{Top  panel: time  series of  the responses  detected  at the
   differential  channel  of  the  interferometer  when  we  inject  4
   different sinusoidal signals in the onboard coils. The amplitude of
   these sinusoids in all the  cases are 1\,mA and the frequencies are
   respectively  0.5\,mHz, 2\,mHz,  5\,mHz and  9\,mHz.  Bottom panel:
   Fourier analysis of  the time series displayed in  the top panel of
   this figure.}
 \label{fig:FreqLoop}
\end{figure}

The SNR  across the instrument  measurement bandwidth for each  of the
channels --- $o_{x_{12}}$, $o_{\eta_1}$, and $o_{\phi_1}$ --- is shown
in  Fig.~\ref{fig:SNR}.  The SNR  reaches its  maximum between  0.5 to
1.5\,mHz for  the displacement reading, and  from 1 to  2\,mHz for the
rotation channels.  This is the most sensitive band of the instrument.
This is confirmed by inspecting Fig.~\ref{fig:FreqLoop}, where we show
the response of the system to the excitation by 4 different sinusoidal
currents.   All these  sinusoidal  currents have  the same  amplitude,
1\,mA, but they oscillate respectively at 0.5\,mHz, 2\,mHz, 5\,mHz and
9\,mHz. In  the top  panel, we show  the readings of  the differential
displacement channel to this set  of four sinusoids.  When exciting at
0.5\,mHz the  amplitude is  $\sim 40$~nm, whereas  at 2\,mHz  drops to
$\sim 5$~nm.  Finally,  when the frequency is 5\,mHz  the amplitude of
the excursion is only $\sim 1$~nm. This same effect is observed in the
bottom panel, where we show the Fourier analysis of these time series.
As can  be seen,  each of  the readings has  a frequency  component at
$\omega_0$ and a second one at $2\omega_0$, as expected.  Note as well
that the $2\omega_0$ components  are highly attenuated with respect to
the  main component because  they are  located at  higher frequencies.
This simple  analysis seems to indicate that  the excitation frequency
should be chosen around 1\,mHz.  However, this range of frequencies is
where  the uncertainty  of  the capacitive  actuator  has the  largest
impact  on   the  estimates  of   the  magnetic  parameters   ---  see
Fig.~\ref{fig:sensitivity}  and section  \ref{sec:sensitivity}.  Thus,
the determination  of the optimal  excitation frequency should  be the
result of a joint optimization procedure, taking into account both the
frequency dependence of  the SNR and the uncertainties  in the gain of
the capacitive actuator.

\begin{figure}[!t]
 \centering
 \includegraphics[width = \columnwidth]{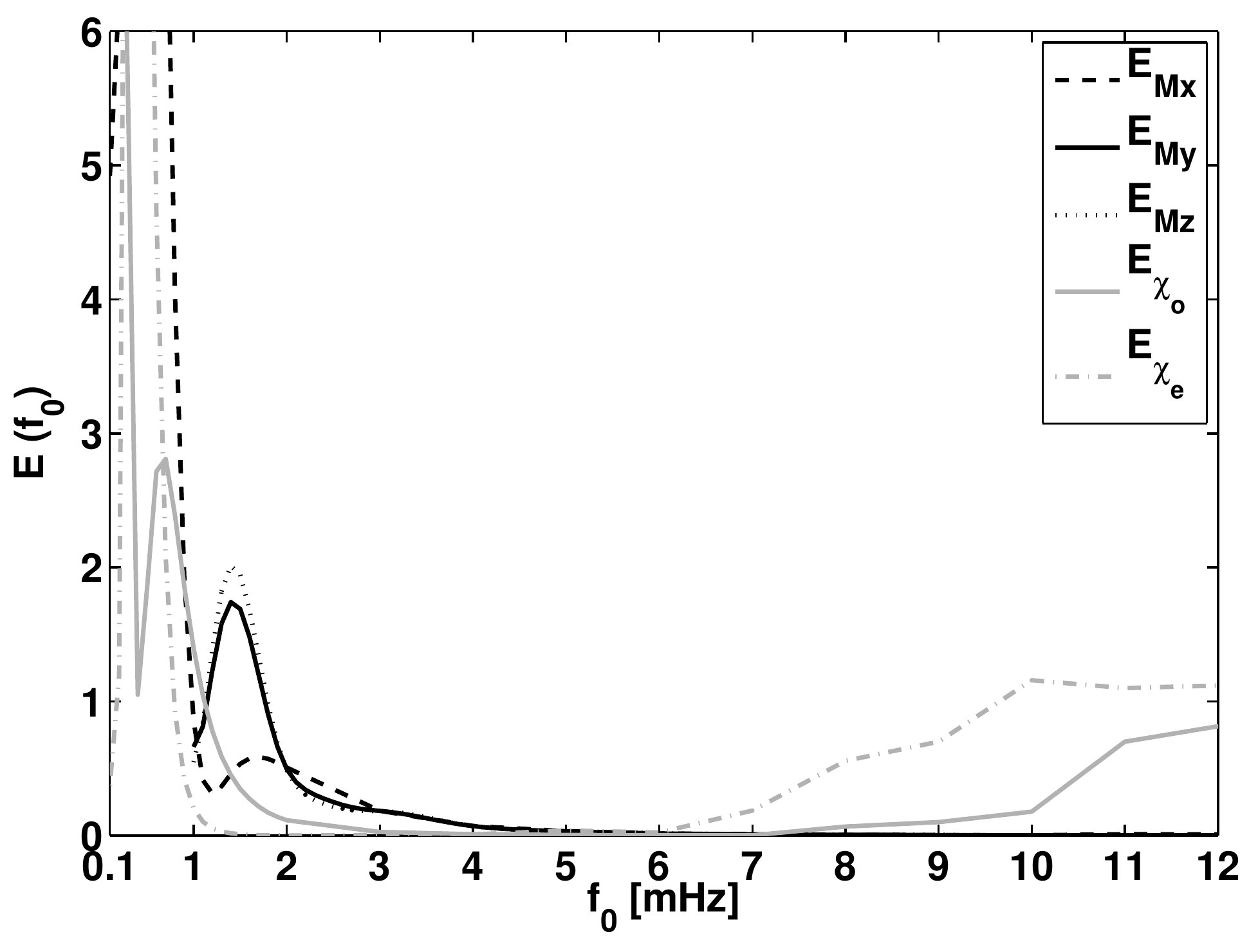}
 \caption{Global error  functions of  each of the  magnetic parameters
   with respect to the excitation frequency.  The dashed black line is
   the global error function for $M_x$ --- that is, $E_{M_x}(f_0)$ ---
   the black solid  line is $E_{M_y}(f_0)$, and the  dotted black line
   corresponds  to  $E_{M_z}(f_0)$.   Finally,  the  solid  gray  line
   corresponds to $E_{\chi_{\rm o}}(f_0)$  and the dashed gray line to
   $E_{\chi_e}(f_0)$.}
\label{fig:OptimFreq}
\end{figure}

To  find the optimum  excitation frequency  we compute  the estimation
error of  each of the magnetic parameters  for different uncertainties
of the gain of the capacitive  actuator ranging from $-0.2$ to 0.2. We
do  this  for  different  excitation  frequencies  across  the  entire
measurement bandwidth.  Thus, for  each magnetic parameter, we compute
an error function for  each gain uncertainty, $e_ b(\omega_0)$.  Then
we add  quadratically each of  these functions with  their appropriate
weight factor:

\begin{equation}
 E(\omega_0) = \sum_b \left(\frac{1}{b} e_b(\omega_0) \right)^2
 \label{eq.totalerror}
\end{equation}

\noindent  where $b$  is the  uncertainty in  the  capacitive actuator
gain, in  percentage. In this way  we compute a  global error function
for   each   of    the   magnetic   parameters,   $E_{M_x}(\omega_0)$,
$E_{M_y}(\omega_0)$,  $E_{M_z}(\omega_0)$, $E_{\chi_o}(\omega_0)$, and
$E_{\chi_e}(\omega_0)$.   The  absolute   minima  of  these  functions
correspond  to  the  best  excitation  frequencies  for  each  of  the
parameters.

\begin{figure*}[!t]
 \centering
 \includegraphics[width = 0.7\textwidth]{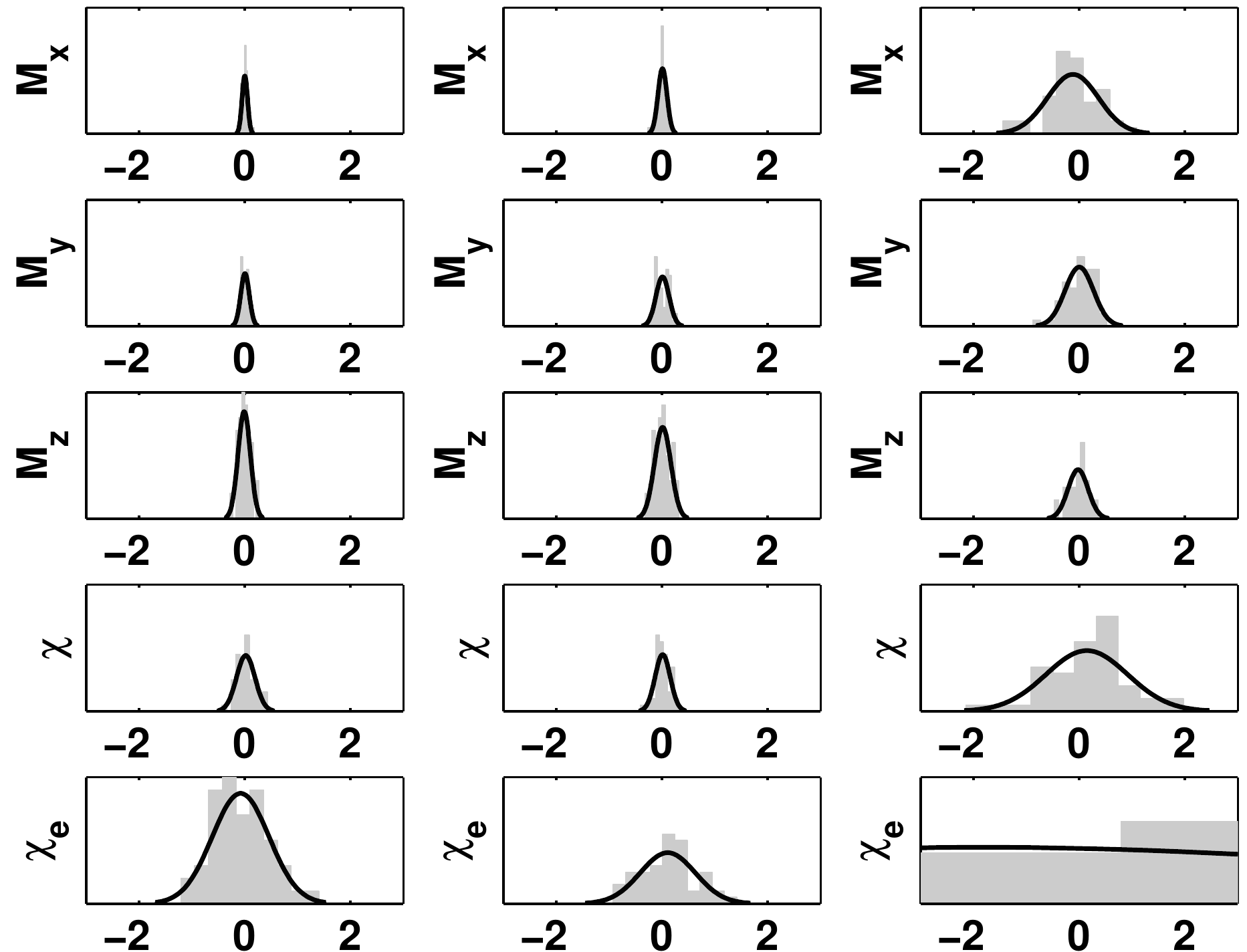}
 \caption{Statistical  distributions  of   the  estimates  for  the  5
   magnetic   parameters  of   the   test  masses   for  3   different
   simulations. The left column  shows the statistical distribution of
   these  parameters   when  the  capacitive  actuator   gain  has  no
   uncertainty. This simulation is done for an excitation frequency of
   5\,mHz. The second column  shows the results when the uncertainties
   of the capacitive actuator gains  of the principal axes are modeled
   with  a  normal  distribution   of  zero  mean  and  0.01  standard
   deviation. This  simulation is done for an  excitation frequency of
   5\,mHz,  too.   Finally,  the  third column,  shows  the  parameter
   estimation results for the same experiment as in the second column,
   but for  1\,mHz.  The $x$-axis  of each subplot shows  the relative
   error in the parameter (in percentage).}
 \label{fig:Results}
\end{figure*}

The  global  error  functions  computed  in  this  way  are  shown  in
Fig.~\ref{fig:OptimFreq} for frequencies from 0.1\,mHz to 12\,mHz. For
the case  of remanent  magnetic moment the  error function  presents a
very broad minimum  between $\sim 5$~mHz and $\sim  11$~mHz, being the
absolute minimum at $\sim 10$~mHz.  Note that at lower frequencies the
global  error  function grows  very  abruptly.   This occurs  because,
although the  SNR of  the experiment is  larger at  these frequencies,
they are also very sensitive  to the biases introduced by the actuator
uncertainty.  Note as well that  the error functions have local minima
at  around 1\,mHz, and  also a  local maximum  between $\sim  1.2$ and
$\sim  2$~mHz,  following  the  sensitivity curve  of  the  capacitive
actuator ---  see Fig.~\ref{fig:sensitivity}.  On the  other hand, the
optimal frequency needed to estimate $\chi_{\rm o}$ and $\chi_{\rm e}$
lies between 5  and 7\,mHz. This is because  these last two parameters
are estimated  with the $2\omega_0$ component of  the $x_{12}$ output,
and higher frequencies are penalized by the larger attenuation on this
frequency component.   Finally, it is worth mentioning  that the phase
shift shown in  Fig.~\ref{fig:sensitivity} around 1\,mHz penalizes the
estimation at  low frequencies, because the  components at $2\omega_0$
suffer a  different and unknown  shift with respect to  the $\omega_0$
component.   In summary, the  best choice  of excitation  frequency is
5\,mHz to  estimate $\chi_{\rm o}$  and $\chi_{\rm e}$ and  10\,mHz to
estimate the  three components of the  magnetic moment.  Nevertheless,
if only  one inflight  experiment could be  performed due  to planning
restrictions of the mission, the  best frequency would be 5\,mHz. This
value is the result of minimizing  the quadratic sum of the five error
functions of the five parameters.


\section{Robustness of the estimates}
\label{chap.7}

Finally, to conclude with our  analysis we have studied the robustness
of our findings.  Specifically, we  have tested the performance of our
estimation algorithm  under several  circumstances. In order  to model
statistically  its   performance,  we  have   estimated  the  magnetic
parameters for  50 different simulated experiments  and calculated the
statistical distribution of the relative errors of each parameter. For
example, for the case of $M_x$ the relative error is computed as:

\begin{equation}
e_{M_x} = \frac{(\hat{M}_x-M_x)}{M_x}
\end{equation}

\noindent  where  $\hat{M}_x$ is  the  estimated  parameter and  $M_x$
represents its  true value.  

Here we  present the results  of three different simulations.   In the
first simulation  we excite the coils  with a 5 \,mHz  sinusoid and we
consider that the gain of  the capacitive actuator is the nominal one.
In the second simulation  we maintain the 5\,mHz excitation frequency,
but in  this case the gains  of the capacitive actuators  of the three
main axis are  modeled with a normal random  distribution of zero mean
and of 0.01 standard deviation.  Finally, the third simulation is only
performed for illustrative purposes. We maintain a random distribution
of  the uncertainty of  the gain  of the  capacitive actuator,  but we
excite  the  coils at  1  \,mHz.  Note  that the  analysis  previously
explained in  Sect.~\ref{chap.4} concluded that  this frequency should
not be  used.  Consequently, this case clearly  illustrates the effect
of choosing a wrong excitation frequency.

The  error  distributions  for  each  of  the  simulations  previously
described and  for each  of the magnetic  parameters are  displayed in
Fig.~\ref{fig:Results}, and  their respective standard  deviations are
listed  in  Table   \ref{tab:tableresults}.   For  consistency,  these
results are checked against  the Cram\'er-Rao lower bound, which gives
a    lower    limit    for    the   variance    of    the    estimated
parameters~\cite{bib:leastSquares}.  The Cram\'er-Rao bounds  for each
of the estimates  of the magnetic parameters are  listed the first row
of   Table  \ref{tab:tableresults}.    Moreover,  for   each   of  the
simulations presented here  we also compute the ratio  of the standard
deviation to the  Cram\'er Rao lower bound.  For  the first simulation
we obtain variances close to the Cram\'er-Rao lower bound, as expected
due to  the large SNR.  In  the second numerical  experiment we obtain
standard  deviations   smaller  than  0.18\%  for   all  the  magnetic
parameters,  except for  $\chi_{\rm e}$,  which  is the  one with  the
lowest SNR.   In this  experiment, we are  still close to  the optimal
Cram\'er Rao  bound because we  minimize the effect of  the capacitive
actuator uncertainty.   Finally for the third simulation  we obtain an
important degradation  of the performance of  the parameter estimation
procedure.  In particular,  the standard  deviations are  increased by
more  than  1  order of  magnitude.  The  ratio  with respect  to  the
Cram\'er-Rao lower  bound is also clearly  much larger.  Particularly,
the  performance  of  the   estimate  of  $\chi_{\rm  e}$  is  totally
unacceptable for this  experiment, obtaining an estimation performance
~25 times  worse than the  optimal one. Finally, comparing  the second
and  third columns of  Fig.~\ref{fig:Results} ---  and the  second and
third sections of Table~\ref{tab:tableresults} --- we confirm that our
estimation procedure  delivers better  results (and close  to optimal)
for an excitation  frequency of 5\,mHz than for  1\,mHz, which was the
frequency adopted  in the preliminary  design of the  experiment. This
clearly  demonstrates  the  importance  of  choosing  the  appropriate
excitation frequency.

\begin{table}[t!]
 \caption{Standard   deviations  of   the  estimated   parameters  for
   different  estimation   scenarios.   For  each   of  the  different
   scenarios we calculate the ratio between the actual performance and
   the optimal Cram\'er Rao lower bound.}
 \label{tab:tableresults}
\begin{ruledtabular}
\begin{tabular}{crcccccc}
 Run   &  & $\Delta\hat{M}_x$ & 
            $\Delta\hat{M}_y$ & 
            $\Delta\hat{M}_z$ &
            $\Delta\hat{\chi_{\rm o}}$ & 
            $\Delta\hat{\chi_{\rm e}}$ \\ 
\hline 
& CR bound                    \quad  & 0.019\% & 0.046\% & 0.139\% & 0.083\%  & 0.263\% \\
\hline 
\multirow{2}{*}{1} & $\sigma$ \qquad & 0.028\% & 0.067\% & 0.156\% & 0.084\%  & 0.557\% \\
 & CR ratio                   \quad  & 1.47    & 1.45    & 1.12    & 1.01     & 2.11    \\
\hline 
\multirow{2}{*}{2} & $\sigma$ \qquad & 0.123\% & 0.132\% & 0.162\% & 0.176\%  & 0.632\% \\
& CR ratio                    \quad  & 6.47    & 2.86    & 1.17    & 2.12     & 2.40    \\
\hline 
\multirow{2}{*}{3} & $\sigma$ \qquad & 0.331\% & 0.215\% & 0.445\% & 0.553\%  & 6.557\% \\
& CR ratio                    \quad  & 17.42   & 4.67    & 3.20    & 6.66     & 24.93   \\
\end{tabular}
\end{ruledtabular}
\end{table}

However, this  is not the most  robust estimate that  can be obtained.
In  particular,  we  suggest   to  use  a  multi-frequency  estimation
technique, where the properties of  the test masses are computed using
the  results  obtained at  different  frequencies.   In  this way  the
effects  of  spurious  or  non-modeled  effects at  a  given  specific
frequency can  be minimized.  This  can be done weighting  the results
obtained for each of the  magnetic parameters at each frequency by the
inverse   of  the   corresponding  total   error  function   given  by
Eq.~(\ref{eq.totalerror}).  For instance, for the $x$-component of the
remanent magnetic moment we may write:

\begin{equation}
\hat{M}_x = \sum_{i=1}^{N}\frac{1}{E_{M_x}(\omega_i)} \, \hat{M}_{x_{\omega_i}}
\end{equation}

\noindent  where  $N$  is   the  total  number  of  frequencies  used,
$\omega_i$     is    the    corresponding     excitation    frequency,
$\hat{M}_{x_{\omega_i}}$ is  the estimate  of $M_x$ at  $\omega_i$ and
$\hat{M}_x$  is  the  final   combined  estimate.   In  this  equation
$E_{M_x}(\omega_i)$      are     the     weighting      factors     of
Eq.~(\ref{eq.totalerror}) adequately normalized:

\begin{equation}
\sum_{1=1}^N\frac{1}{E_{M_x}(\omega_i)} = 1.
\end{equation}

This  estimation  procedure  provides  an  estimate  of  the  magnetic
characteristics of  the test  masses that takes  into account  all the
limiting factors of the  LTP instrument, and also delivers estimations
which  are robust  to  other unexpected  (and  not modeled)  frequency
dependent effects.

\begin{table*}[t]
\caption[Uncertainty   of   the   determination   of   each   magnetic
  contribution.]{Budget  of  the contribution  of  the magnetic  field
  effects    to    the   total    acceleration    noise   and    their
  uncertainties. These  uncertainties are computed  using the expected
  error  of  the  magnetic  characteristics reported in this paper and using the expected error of the magnetic field
  determination reported in Refs. \cite{bib:myPaper, bib:myPaper2}}.
\label{tab2}
\begin{center}
\begin{tabular}{l l }
\hline\hline
Contribution                                   		&  \qquad  Differential acceleration noise [m s$^{-2}$ Hz$^{-1/2}$] \\
 \hline
  Fluctuation of the spacecraft magnetic field     	&  \qquad (0.680 $\pm \, 0.096$) $\times 10^{-15}$ \\
  Fluctuation of the spacecraft magnetic field gradient &  \qquad (1.097 $\pm \, 0.108$) $\times 10^{-15}$ \\
  Down converted AC magnetic fields      		&  \qquad (1.265 $\pm \, 0.254$) $\times 10^{-15}$\\
  Interplanetary magnetic field fluctuation  		&  \qquad (1.701 $\pm \, 0.241$) $\times 10^{-15}$\\
  Lorentz force						&  \qquad (0.013 $\pm \, 0.001$) $\times 10^{-15}$\\
\hline
Total                                                   &  \qquad (2.775 $\pm \, 0.425$) $\times 10^{-15}$\\
\hline
Requirement                                             &  \qquad 12.0 $\times 10^{-15}$ \\
  \hline\hline
\end{tabular}
\end{center}
\end{table*}


\section{Magnetic contribution to proof-mass acceleration noise}
\label{chap.MagCont}

In previous sections we have shown that the magnetic properties of the
test masses can  be characterized in flight with  accuracies below 1\%
for  each of  the parameters.   The ultimate  goal of  measuring these
properties is  the determination of  the magnetic contribution  to the
proof-mass acceleration  noise.  To do  this, the estimated  values of
the magnetic field and magnetic field gradient at the positions of the
test  masses are  needed.   It has  been concluded  \cite{bib:myPaper,
bib:myPaper2}  that, in  the worst  case, the  magnetic field  and its
gradient can be estimated with  an accuracy better than 10\%. The spatial nonhomogeneities of the field are included in this uncertainty. Thus, an
important goal  consists in assessing how  these uncertainties project
into the precision of our estimate of the magnetic contribution to the
total  differential  acceleration  reading  in LPF.   Accordingly,  we
propagate the errors of  the magnetic characteristics and the magnetic
field and  gradient into the calculation of  the magnetic acceleration
noise:

\begin{equation}
\sigma_{\rm total}    =   \sqrt{\sum_{i=1}^N \left(\frac{\partial f}
{\partial s_i} \sigma_{s_i} \right)^2},
\label{eq.errors}
\end{equation}

\noindent  where  $f$  is  the  total  magnetic  contribution  to  the
acceleration noise and $s_i$ are the several sources of error.  In our
case,  the  total  magnetic  acceleration  noise is  the  sum  of  the
fluctuation of the  magnetic field and magnetic field  gradient of the
spacecraft, the down converted  AC magnetic fields, the interplanetary
magnetic  field  fluctuations,  and  the Lorentz  force  contributions
\cite{bib:EPF}, whereas the sources  of error are the uncertainties of
the magnetic field and its gradient and those of the remanent magnetic
moment and  susceptibility of the  test masses.  The  results obtained
using Eq.~(\ref{eq.errors})  and the computed  uncertainties are shown
in Table~\ref{tab2}. The total  magnetic contribution, which is $2.775
\times  10^{-15}$~m~s$^{-2}$~Hz$^{-1/2}$ is  determined  with with  an
accuracy of $0.425 \times 10^{-15}$~m~s$^{-2}$~Hz$^{-1/2}$. This means
that the  magnetic contribution  to the total  noise can  be estimated
with a fair  accuracy, and therefore, can be  subtracted from the main
acceleration reading, with a relative  error 15\% across the whole LTP
measurement bandwidth.  This represents an enhancement of one order of
magnitude with respect to previous studies.


\section{Summary and conclusions}
\label{chap.8}

In this paper we have  studied how the magnetic characteristics of the
test  masses  onboard LISA  Pathfinder  can  be  determined.  This  is
essential to  estimate the magnetic  noise contribution to  the entire
noise budget  and, most importantly,  to subtract this noise  from the
displacement  reading.    The  estimation  of   $M_x$,  $M_y$,  $M_z$,
$\chi_{\rm o}$  and $\chi_{\rm e}$  is done by injecting  a controlled
magnetic  field at  the position  of the  test masses.   The  field is
generated by a sinusoidal  current circulating through the two onboard
induction coils placed at each  side of both test masses.  The induced
magnetic  field results  in magnetic  forces and  torques on  the test
masses  that excite  their dynamics.   We  have shown  that the  force
acting  on the  test masses  has  two frequencies,  while the  torques
oscillate at single frequency,  allowing to estimate the properties of
the test  masses by  an adequate processing  of three of  the readings
delivered by the interferometer.   These readings are the differential
displacement of both test  masses ($o_{x_{12}}$), the rotation of test
mass 1 about  the $y$-axis ($o_{\eta_1}$) and that  about the $z$-axis
($o_{\phi_1}$).  We have also shown that the time series received from
the satellite's  telemetry need to  be whitened and split  to minimize
the   low-frequency  effects   inherent  in   the  operation   of  the
instrument. This  way, the magnetic  parameters can be estimated  by a
classical  single-channel  least-square  technique  once  the  effects
produced by cross-talks are determined and corrected. Additionally, we
have  assessed the  sensitivity  of the  estimation  procedure to  the
uncertainty in the gain of  the capacitive actuator of the instrument.
This effect showed  to be very relevant and,  most importantly, it has
been found to depend on the excitation frequency. Moreover, the SNR of
the received  signals also  depends on the  frequency of  the injected
signal.   Accordingly, we  have  also presented  a joint  optimization
analysis that  takes into  account these two  factors, leading  to the
conclusion  that   the  optimal  excitation  frequency   for  a  joint
experiment  is 5\,mHz.   Performing the  experiment at  this frequency
allows to estimate the magnetic characteristics without being affected
by the  likely uncertainty in  the capacitive actuator gain.   In this
case we obtain parameter variances  smaller than $\sim 0.7$\% when the
deviations  of the  gain of  the capacitive  actuator are  $\sim 1$\%.
Using all the previously  explained steps and adopting this excitation
frequency,  the estimation  turns out  to be  more accurate  than that
obtained using the  preliminary design of the experiment,  for which a
frequency of 1\,mHz  was adopted.  Moreover, we have  suggested that a
multi-frequency  estimation technique could  deliver estimates  of the
highest quality,  enhancing the robustness of the  experiment in front
of  non-modeled  frequency-dependent  effects.  Finally,  taking  into
account the  aforementioned results,  together with the  results about
magnetic field  estimation presented  elsewhere, we estimate  that the
total magnetic  contribution to the proof-mass  acceleration noise can
be  determined with a  $\sim 15\%$  error level  across the  whole LTP
frequency band.


\begin{acknowledgments}
This work  was partially supported  by MCINN grants  ESP2007-61712 and
AYA08--04211--C02--01.  Part  of this work  was also supported  by the
AGAUR.
\end{acknowledgments}
  


\end{document}